\def\bq{\begin{equation}}
\def\eq{\end{equation}}
\def\bqa{\begin{eqnarray}}
\def\eqa{\end{eqnarray}}
\def\bqb{\begin{eqnarray*}}
\def\eqb{\end{eqnarray*}}
\documentstyle[12pt]{article}\textwidth 16cm
\def\pr#1#2#3{ Phys. Rev. ${\bf{#1}}$ (#2) #3}

\def\pl#1#2#3{ Phys. Lett. ${\bf{#1}}$ (#2) #3 }

\def\np#1#2#3{ Nucl. Phys. ${\bf{#1}}$ (#2) #3}
\def\zp#1#2#3{ Z. Phys. ${\bf{#1}}$ (#2) #3}

\def\eg{{\it e.g.\/}}

\global\nulldelimiterspace = 0pt

\def\Bsl{\hbox{/\kern-.6700em$B$}} 
\def\Dsl{\hbox{/\kern-.6700em$D$}} 
\def\Wsl{\hbox{/\kern-.6700em$W$}} 
\def\roughly#1{\mathrel{\raise.3ex
    \hbox{$#1$\kern-.75em\lower1ex\hbox{$\sim$}}}}

\def\ol#1{\overline{#1}}
\def\L{ {\cal L }}
\def\O{ {\cal O }}
\def\sw{s_W}
\def\cw{c_W}
\def\swd{s^2_W}
\def\cwd{c^2_W}

\def\mwd{M_W^2}
\def\mw{M_W}
\def\lw{\lambda_W}
\def\rd{\sqrt2}
\def\ep#1#2{(\epsilon_{#1}\epsilon_{#2})}
\def\dh#1{ {1\over D_H({#1})} }
\def\nh#1{  D_H({#1}) }
\def\co{\biggm[}
\def\cf{\biggm]}
\begin{document}
\pagenumbering{arabic}
\thispagestyle{empty}

\begin{flushright}
 PM/95-11 \\ THES-TP 95/06 \\
May 1995 \end{flushright}
\vspace{2cm}
\begin{center}
{\Large\bf Signatures of anomalous couplings in boson pair
production through $\gamma \gamma$ collisions}
\footnote{Partially supported by the EC contract CHRX-CT94-0579.}
 \vspace{1.5cm}  \\
{\large G.J. Gounaris$^{a}$, J.Layssac$^b$ and F.M. Renard$^b$}
\vspace {0.5cm}  \\
$^a$Department of Theoretical Physics, University of Thessaloniki,\\
Gr-54006, Thessaloniki, Greece,\\
\vspace{0.2cm}
$^b$Physique
Math\'{e}matique et Th\'{e}orique,
CNRS-URA 768,\\
Universit\'{e} de Montpellier II,
 F-34095 Montpellier Cedex 5.\\

\vspace {2.0cm}

 {\bf Abstract}
\end{center}

\noindent
We discuss possible New Physics (NP) effects on the processes
$\gamma\gamma \to W^+W^-$, $ZZ$,
$Z\gamma$, $\gamma\gamma$, $HH$ which are observable in
$\gamma\gamma$ collisions. Such collisions may be achieved
through laser backscattering at a high
energy $e^+e^-$ linear collider. To the extent that no new
particles will be directly produced in the future
colliders, it has already been
emphasized that the new physics possibly hidden in the
bosonic interactions, may be represented by
the seven $dim=6$ operators $\O_W$, $\O_{B\Phi}$, $\O_{W\Phi}$,
$\O_{UB}$, $\O_{UW}$, $\ol{\O}_{UB}$
and $\ol{\O}_{UW}$ (the last two ones being CP-violating).
In this paper, we show that the above processes are
sensitive to NP scales at the several TeV range, and we subsequently
discuss the possibility to disentangle the effects of the
various operators.

\clearpage

\section{Introduction}
The search for manifestations of New Physics (NP) is part of the
program of future high energy colliders \cite{coll, phys}. As discussed in
previous papers \cite{DeR, blind, GRVZbb}, if no new particles are
produced in the future colliders, these NP manifestations
may appear only as anomalous interactions among the particles
already present in SM. In this paper we
restrict to the study
of the purely bosonic part of such forms of NP \cite{DeR, blind,
model}. Within a framework like this, NP effects have
been searched
for by using the high precision measurements obtained at
LEP1 \cite{lep1}. High energy $e^+e^-$ linear colliders will offer many
more possibilities though, to test the sector of the gauge
boson and Higgs interactions with a high
accuracy. The most famous process is $e^+e^- \to W^+W^-$
\cite{Gaemers}. This has
been carefully studied and it has been shown that indeed the 3-gauge
boson vertices $\gamma W^+W^-$ and $Z W^+W^-$ can be very
accurately constrained through it \cite{BMT, Bilenky}.\par

 Another type of processes accessible at a
linear collider is boson-boson scattering \cite{bb}. Such
processes are even more interesting, since they are
sensitive not only to the 3-gauge boson vertices, but also to
4-gauge boson \cite{Schrempp, Yeduhai, Cheung} as well as to
Higgs couplings \cite{higpro}. Thus the scalar (Higgs)
sector, which is presently the most mysterious part of the
electroweak interactions, may be tested in a much deeper
way using the $\gamma\gamma$ collisions offered by the laser
backscattering method \cite{laser}.\par

In a $\gamma \gamma $ collider, the five boson pair
production processes available to be studied are $\gamma\gamma \to
W^+W^-$, $ZZ$, $Z\gamma$, $\gamma\gamma$ and $HH$
(assuming that the physical Higgs particle exists and it is
not very heavy). The purpose of the present work is to show that
the study of the $p_T$ distribution of one
of the final bosons provides a very sensitive test of the
various possible forms of NP. At present we leave aside the process
$\gamma\gamma \to H$, which indeed provides a very powerful way to
study the anomalous couplings of the Higgs particle. This
single Higgs production process has been first considered in
\cite{higpro} and is thoroughly trated in the companion paper
\cite{higpro1}.
In the present paper we concentrate on the boson pair
production due to both gauge and
Higgs boson exchanges. The specific illustrations we give assume a
Higgs mass in the 100 GeV region.\par

Assuming that the NP scale $\Lambda_{NP}$ is sufficiently large,
the effective NP Lagrangian is satisfactorily
described for our purposes in terms of
$dim=6$ bosonic operators only \cite{Leff, blind}.
There exist only seven $SU(2)\times U(1)$ gauge
invariant such operators called "blind",
which are not strongly constrained by existing LEP1
experiments \cite{DeR, blind, model}. Three of them
\bqa
\O_W &= & {1\over3!}\left( \overrightarrow{W}^{\ \ \nu}_\mu\times
  \overrightarrow{W}^{\ \ \lambda}_\nu \right) \cdot
  \overrightarrow{W}^{\ \ \mu}_\lambda \ \ \
   \ \ , \\[0.1cm]
\O_{W\Phi} & = & i\, (D_\mu \Phi)^\dagger \overrightarrow \tau
\cdot \overrightarrow W^{\mu \nu} (D_\nu \Phi) \ \ \  , \ \
 \\[0.1cm]
\O_{B\Phi} & = & i\, (D_\mu \Phi)^\dagger B^{\mu \nu} (D_\nu
\Phi)\ \ \  , \
\eqa
induce anomalous triple gauge boson couplings, while
the remaining four\footnote{In
the definition of $\O_{UW}$ and $\O_{UB}$ we have ~subtracted a
trivial contribution to the $W$ and $B$ kinetic energy respectively.}
\bqa
\O_{UW} & = & \frac{1}{v^2}\, (\Phi^\dagger \Phi - \frac{v^2}{2})
\, \overrightarrow W^{\mu\nu} \cdot \overrightarrow W_{\mu\nu} \ \
\  ,  \ \ \\[0.1cm]
\O_{UB} & = & \frac{4}{v^2}~ (\Phi^\dagger \Phi -\frac{v^2}{2})
B^{\mu\nu} \
B_{\mu\nu} \ \ \  , \ \ \\[0.1cm]
 \ol{\O}_{UW} & = & \frac{1}{v^2}\, (\Phi^\dagger
\Phi)
\, \overrightarrow W^{\mu\nu} \cdot
\widetilde{\overrightarrow W}_{\mu\nu} \ \
\  ,  \ \ \\[0.1cm]
\ol{\O}_{UB} & = & \frac{4}{v^2}~ (\Phi^\dagger \Phi )
B^{\mu\nu} \
\widetilde{B}_{\mu\nu} \ \ \  \ \ \
\eqa
create  anomalous CP conserving and CP violating
Higgs couplings. These later four operators constitute a
dedicated probe of the scalar sector.
Ideas on how these operators could be generated in various
NP dynamical scenarios have been discussed in \cite{model, dyn}.
The effective Lagrangian describing the NP induced by these
operators is given by
\bqa
\L_{NP} & = & \lw \frac{g}{\mwd}\O_W +
{f_B g\prime \over {2M^2_W}}\O_{B\Phi} +{f_W g \over
{2M^2_W}}\O_{W\Phi}\ \ + \nonumber \\
\null & \null &
d\ \O_{UW} + {d_B\over4}\ \O_{UB} +\ol{d}\ \ol{\O}_{UW} +
{\ol{d}_B\over4}\ \ol{\O}_{UB} \ \ \ \
\ \ ,  \
\eqa
which fully defines the various couplings.
Quantitative relations between these
couplings and the corresponding NP scales have been established on the
basis of the unitarity constraints \cite{unit, uni2, model}. \par

Below, we first write the amplitudes for the five processes
mentioned above, in terms of
the seven couplings associated to the operators considered.
We then discuss procedures
towards disentangling the effects of these operators.
This allows us to express the observability limits for the
various forms of new physics, in terms
of the related NP scales.  We find that this can be achieved to a
large extent by just using only the aforementioned $p_T$
distribution. Only for the disentangling of the
CP violating operators it is
necessary to augment this simple analysis by also looking
at the density matrix of the final $W$ or $Z$ bosons.\par

In Sect. 2 we recall the formulation of $\gamma \gamma$
collisions through the laser backscattering method and give the
expressions for the luminosities and the invariant mass and $p_T$
distributions. These distributions are expressed in terms of
helicity amplitudes most of which are computed in
\cite{LHCGLR, higpro, uni2}. In Appendix A we just give the analytic
expressions for their NP contributions in the high energy limit.
Sect. 3 contains an overview of the characteristics of each of
the five $\gamma \gamma $ processes, emphasizing their
dependence on the various anomalous couplings. The precise
sensitivity to each operator is discussed in Sect. 4 and the
possible ways to disentangle them in Sect. 5. Concluding words
are given in Sect. 6.\par

\section{Laser induced $\gamma\gamma$ collisions}

The boson pair production processes that we shall consider here are
$\gamma \gamma \to W^+W^-$,  $\gamma \gamma \to ZZ$,
$\gamma \gamma \to \gamma Z$,
$\gamma \gamma \to \gamma \gamma$, and also $\gamma \gamma \to HH$.
\par
In laser induced $\gamma \gamma$ collisions at $e^+e^-$
colliders, with unpolarized $e^\pm$ and laser beams,
photon fluxes are given in terms of the photon distribution
\bq
 f^{laser}_{{\gamma /e}}(x)~ =~{1\over{D(\xi)}}
\left(1-x+{1\over{1-x}}-{4x\over{\xi(1-x)}}
+{4x^2\over{\xi^2(1-x)^2}}\right)
  \ \ \ ,\ \
\eq
where $x$ is the fraction of the incident $e^{\pm}$ energy carried by
the backscattered photon, and
\bq
  D(\xi) =
\left(1-{4\over{\xi}}-{8\over{\xi^2}} \right) Ln(1+\xi)\, +\,
{1\over2}+{8\over{\xi}}\,-\,
{1\over{2(1+\xi)^2}}
  \ \ \ ,\ \
\eq
with $\xi=2(1+ \sqrt{2}) \simeq 4.8$ ,
 $x_{max}={\xi\over{1+\xi}}\simeq 0.82$
\cite{fgamma,Kuhn}.\par

The invariant mass distribution is obtained as
\bq
{d\sigma \over{dy}}~ =~ {d{\cal L}_{\gamma \gamma}
\over{dy}}~\sigma_{\gamma \gamma}(s_{\gamma \gamma}) \ \ \ \ \
\ ,\ \ \ \
\eq
where
\bq
\frac{d{\cal L}_{\gamma \gamma}(y)}{dy}~ =~ 2y \int ^{x_{max}}
_{\tau\over{x_{max}}}\ {dx\over{x}}\, f^{laser}_{\gamma/e}(x)\,
f^{laser}_
{{\gamma /e}}\left({\tau \over{x}}\right) \ \ \ \ \ \ \ \ \
\ \ \ , \ \ \ \
\eq
and
\bq
y \equiv \sqrt{\tau}~ \equiv ~\sqrt { \frac{s_{\gamma
\gamma}}{s_{ee}}} \ \ \ \ \ \  \ \ \ \ \ \  \eq
is the ratio
of the $\gamma \gamma $ c.m.\@ energy to the $e^+e^-$ one
satisfying
\bq
y < y_{max}~ \equiv ~x_{max}\simeq 0.82 \ \ \ \ . \
\eq
The correspondingly expected number of events per year
is
\bq
{dN\over{dy}}~ =~ \bar {\cal L}_{ee}\, {d\sigma \over{dy}} \
\ \ \ \ \ ,\ \ \ \ \ \ \
\eq
where $\bar{\cal L}_{ee}$ is the integrated $e^+e^-$ annual
luminosity taken to be $20, 80, 320 fb^{-1} year^{-1}$ for a
0.5, 1. or 2. TeV collider respectively.\par

Of course, forward boson production
should be cut-off in a realistic situation, since these
events are ~inevitably lost along the beam-pipe. An efficient way
of doing this is by looking at the transverse-momentum
distribution of one of the final bosons $B_3$ and $B_4$ and
cutting-off the small $p_T$ values. For a fixed invariant mass
of the final boson pair, this is given by
\bq
{d\sigma \over{dp_Tdy}}~ =~{y p_T\over {8\pi
s_{\gamma\gamma}|\Delta|}}
 \int ^{x_{max}}
_{\tau\over{x_{max}}}\ dxf^{laser}_{\gamma/e}(x)\,
f^{laser}_
{{\gamma /e}}\left({\tau \over{x}}\right) \Sigma|F(\gamma
\gamma \to B_3 B_4|^2 \ ,
\eq
where
\bqa
|\Delta| & =& {1\over2}\sqrt{s_{\gamma\gamma}(s_{\gamma\gamma}-
4(p^2_T+m^2))} \ \ , \\
& \null &
  p^2_T~<~{s_{\gamma\gamma}\over4}-m^2 \ \ ,
\eqa
and $F(\gamma \gamma \to B_3 B_4)$ is the invariant amplitude
of the subprocess.
Integrating over the invariant mass of the $B_3B_4$ pair we get
\bq
{d\sigma \over{dp_T}}~ =~ \int ^{y_{max}}_{y_0}\
dy{d\sigma \over{dp_Tdy}} \ , \
\eq
with
\bq
y_0~\equiv~{\sqrt{{s_{\gamma\gamma}\over4}-m^2}\over\sqrt{s_{ee}}}
\ \  . \
\eq
This ${d\sigma \over{dp_T}}$ distribution provides a very
useful way for searching for NP, since it not only takes care of
the events lost along the beam pipe, but also because its
measurement does not require full reconstruction of both final bosons.
For the illustrations below we choose
$p_T > p^{min}_T = 0.1 TeV/c$.\par

\section{Description of the Boson Pair Production Processes}

In this section we summarize the properties of each of the five
channels and the way they react to the residual NP
lagrangian.\par

a)$\gamma\gamma \to W^+W^-$\par

The SM contribution consists of $W$ exchange diagrams in the $t$ and
$u$ channels involving the $\gamma WW$ vertex and the
$\gamma\gamma WW$ contact term. There is no Higgs exchange at
tree level. The $W^+W^-$ production rate is copious and the $p_T$
distributions are given in Figs. (1a,b-4a,b).
The NP contributions are induced by the anomalous $\gamma WW$ and
$\gamma\gamma WW$ couplings in the case of the operators $\O_W$,
$\O_{B\Phi}$, and $\O_{W\Phi}$, and by the anomalous
$H\gamma\gamma$ vertex induced by
{}~any of $\O_{UW}$, $\O_{UB}$, $\ol{\O}_{UW}$, $\ol{\O}_{UB}$.
The SM $HWW$ coupling is also modified
by these later four operators.\par

b)$\gamma\gamma \to ZZ$

There is no SM contribution at tree level. The 1-loop
contribution is estimated to be about 100 times smaller than the
tree level contribution to $W^+W^-$ \cite{loop}, so that
$\gamma\gamma \to ZZ$ is
extremely sensitive to genuine NP effects. There exist no neutral
pure gauge anomalous couplings contributing to this process;
(they would only appear if higher dimensional operators were
considered). Thus, the NP contribution comes from Higgs exchange
diagrams in the $s$, $t$, $u$ channels induced by anomalous $H\gamma\gamma,
H\gamma Z$ and $HZZ$ couplings. They are ~generated by the
six operators $\O_{B\Phi}$, $\O_{W\Phi}$, $\O_{UB}$, $\O_{UW}$,
$\ol{\O}_{UB}$ and $\ol{\O}_{UW}$.\par

c)$\gamma\gamma \to \gamma\gamma$ and $\gamma Z$\par

For these processes also there is no SM contribution at
the tree level. NP is only generated by
Higgs exchange diagrams due to the four operators $\O_{UB}$,
$\O_{UW}$, $\ol{\O}_{UB}$, $\ol{\O}_{UW}$. No SM couplings can appear
here. All vertices must be anomalous
and at least one must be $H\gamma\gamma$. Thus the other NP
operators $\O_W$, $\O_{B\Phi}$ nd $\O_{W\Phi}$ give no
contribution at the tree level.\par

d)$\gamma\gamma \to HH$\par

Here also there is no SM contribution at the tree
level. NP is generated by  $\gamma$ and $Z$ exchange
diagrams in the $t$ and $u$ channels, and Higgs
exchange in the s-channel, due to anomalous $H\gamma\gamma$
and $H\gamma Z$ couplings generated by the $dim=6$ operators
$\O_{B\Phi}$, $\O_{W\Phi}$,
$\O_{UB}$, $\O_{UW}$, $\ol{\O}_{UB}$
and $\ol{\O}_{UW}$.
It is also worth remarking that the Higgs exchange diagram ~involves
also the SM $HHH$ vertex.\par

\section{Sensitivity to the various operators}

We now discuss the effect of each of the seven operators treated
one by one.\par

a) The operator $\O_W$ plays a special role because it does not
involve scalar fields. It only affects $W^+W^-$ production
through terms linear and quadratic in $\lambda_W$. The
effects of $\O_W$ on the various $\gamma \gamma $ processes
are given in Figs. 1a,b for $e^+e^-$ energies of $0.5$ and
$1.\ TeV$. The
sensitivity (already studied in \cite{ggVV}) is given in
Table 1.\par

b) The operators $\O_{B\Phi}$ and $\O_{W\Phi}$ are
indistinguishable through these $\gamma \gamma $ processes;
(see Fig 2). The amplitudes for $\gamma \gamma \to W^+W^-$
receive both linear and quadratic contributions in $f_B$ or
$f_W$, while the amplitudes for $\gamma \gamma \to ZZ,\
\gamma Z,\  HH\ $ only get quadratic ones. The highest
sensitivity coming from $W^+W^-$ prediction, is given in
Table 1 and it is comparable to the
sensitivity of $\O_W$. \par

c) The  four operators $\O_{UB}$, $\O_{UW}$,
$\ol{\O}_{UB}$, $\ol{\O}_{UW}$ give linear as well as quadratic
contributions to the amplitudes for $\gamma
\gamma \to W^+W^-$, $\gamma \gamma \to ZZ$
and $\gamma \gamma \to  HH$; see Figs. 3,4.
Note that the contributions of these operators to the amplitudes
for  $\gamma \gamma \to \gamma\gamma$ and $\gamma \gamma \to \gamma Z$,
which are usually the least sensitive ones,
are always quadratic. CP-conserving and
CP-violating operators give essentially the same effects in the
invariant mass and $p_T$ distributions, except for the $HH$
case. This is because the amplitudes for vector boson production
containing the linear terms in the anomalous couplings
do not have any appreciable interference with the SM amplitudes.
Finally, we also note also that the
$\O_{UB}$ contribution may be obtained ~from the $\O_{UW}$ one
by multiplying by the factor ${c^2_W / s^2_W}$. \par

Table 1 summarizes the observability limits expected for each
operator by assuming that for $W^+W^-$ production a departure
of 5\% as compared to the SM prediction in the high $p_T$ range,
will be observable.  For $ZZ$ channel, the observability
limit is set by assuming that a signal of the order of $10^{-2}$ times the
SM $W^+W^-$ rate will be observable. Such an assumption
should be reasonable, considering
the designed luminosities given in Section 2. These
obseravbility limits on the anomalous couplings give essentially
lower bounds for the couplings to be observable. Combining these
bounds with the unitarity relations \cite {unit, uni2, model} we
obtain the upper bounds on the related NP scale which are
indicated in parentheses in Table 1.

\begin{center}
\begin{tabular}{|c|c|c|c|c|} \hline
\multicolumn{5}{|c|}{Table 1: Observability limits for the seven
couplings.} \\
\multicolumn{5}{|c|}
{(in parentheses the corresponding scale in TeV is given)}
\\ \hline
\multicolumn{1}{|c|}{$2E_e$ (TeV)} &
  \multicolumn{1}{|c|}{$|\lambda_W|$} &
   \multicolumn{1}{|c|}{$|d|$ or ${\bar {d}}$} &
    \multicolumn{1}{|c|}{$|d_B|$ or  $|\bar{d}_B|$ } &
      \multicolumn{1}{|c|}{$|f_{B,W}|$} \\ \hline
 0.5 & 0.04 (1.7) & 0.1 (2.4) & 0.04 (4.9) & 0.2 (1.8, 1) \\
 1  & 0.01 (3.5)& 0.04 (5.5)& 0.01 (10) & 0.05 (3.5, 2) \\
 2 & 0.003 (6.4)& 0.01 (21) & 0.003 (19) & 0.015 (6.5, 3.6)  \\ \hline
\end{tabular}
\end{center}
\noindent

\section{Disentangling the various operators}

The purpose of this section is to discuss how one could identify
the origin of an anomalous effect detected in one or several of
the considered channels.
{}From the analysis made in Sect.3,4 a classification of the seven
operators into three different groups has appeared:
\begin{itemize}
\item Group 1 contains $\O_W$ which only affects $W^+W^-$.
\item Group 2 contains $\O_{B\Phi}$ and $\O_{W\Phi}$ which also
affects predominantly the $W^+W^-$ channel and in a weaker way
the $ZZ$, $\gamma Z$, $HH$ ones. If the signal in these three channels
is too weak to be observable, the disentangling from $\O_W$ is possible by
looking at the polarization of the produced $W^{\pm}$. In the
$\O_W$ case $W^+W^-$ are produced in (TT) states whereas in the
$\O_{B\Phi}$ and $\O_{W\Phi}$ cases it is mainly (LL). There is no way
to distinguish the contributions of $\O_{B\Phi}$ from that of
$\O_{W\Phi}$ in $\gamma \gamma $ collisios. The discrimination
between these two operators requires the use of other processes,
like for example $e^+e^- \to W^+W^-$ \cite{BMT}.
\item Group 3 contains the four operators $\O_{UB}$, $\O_{UW}$,
$\ol{\O}_{UB}$, $\ol{\O}_{UW}$ which mainly affect
the channels $W^+W^-$, $ZZ$, $HH$ and in a weaker way the
$\gamma\gamma$ and $\gamma Z$ ones. The comparison of the effects
in $W^+W^-$ and in $ZZ$ production should allow to distinguish this group
from the ~groups 1 and 2. The $ZZ$ final state is mainly (LL) in the
group 3, whereas it is (TT) in group 2. The disentangling of
$\O_{UW}$ from $\O_{UB}$ can be done by looking at the ratios of
the related cross sections. We remark that the linear terms
of the $WW$, $ZZ$ and $HH$
amplitudes for $\O_{UB}$ may be obtained from the corresponding
terms for $\O_{UW}$ by multiplying by the factor  $c^2_W/s^2_W$,
whereas for the quadratic terms this factor is $c^4_W/ s^4_W$ in
$\gamma\gamma$ production and $c^2_W/ s^2_W$ in $\gamma
Z$ production.
\end{itemize}

 There is no way to separate the CP-conserving from the
CP-violating terms in these spectra. More detailed spin analyses
are required, like \eg\@ the search for imaginary parts in
final $W$ or $Z$ spin density matrices obseravable through the decay
distributions \cite{CPanal} or the measurement of asymmetries
associated to linear polarizations of the photon beams \cite{Kraemer}.

\section{Conclusions}

We have shown that boson pair production in real $\gamma\gamma$
collisions is an interesting way to search for NP manifestations
in the bosonic sector. The $\gamma\gamma$ luminosities provided
by laser backscattering at linear $e^+e^-$ colliders should
allow to feel NP effects associated to scales up to
$\Lambda_{th}=20 TeV$  for $2E_e=2 TeV$. This can be achieved
by simply measuring final
gauge boson $p_T$ distributions. As no fermionic states are
involved, any departure from SM
predictions would constitute a clear signal for an anomalous
behaviour of the bosonic sector.
A comparison of the effects in the various final
states $WW$, $ZZ$, $\gamma Z$, $\gamma\gamma$ and $HH$ would
already allow a selection among the seven candidate operators
which should describe the NP ~manifestations. Complete
disentangling should be possible by analyzing final spin states,
i.e. separating $W_T(Z_T)$ from $W_L(Z_L)$ states.
Identification of CP violating terms requires full $W$ or $Z$
spin density matrix reconstruction from their decay
distributions or analyses with linearly polarized photon beams.
The ~occurrence of anomalous terms in gauge boson couplings or in
Higgs boson couplings would be of great interest for tracing
back the origin of NP and its basic properties.

\newpage
{\large \bf Appendix A : Contributions to helicity amplitudes}
\def\ep#1#2{(\epsilon_{#1}\epsilon_{#2})}
\def\dh#1{ {1\over D_H({#1})} }
\def\nh#1{  D_H({#1}) }
\def\co{\biggm[}
\def\cf{\biggm]}

\def\mw{M_W}
\def\mwd{M_W^2}
\def\mz{M_Z}
\def\mzd{M_Z^2}
\def\ct{\cos\theta}
\def\ctp{(\cos\theta+1)\ }
\def\ctm{(\cos\theta-1)\ }
\def\cw{c_W}
\def\sw{s_W}
\def\cwd{c_W^2}
\def\swd{s_W^2}
\def\cs{c_W s_W}
\def\rs{\sqrt s}
\def\rd{\sqrt2}
\def\g{\gamma}
\def\fbd{g^2_1\ f^2_B}
\def\fwd{g^2_2\ f^2_W}
\def\feb{e\ g_1 f_B}
\def\few{e\ g_2 f_W}
\def\fbw{g_1 f_B\ g_2 f_W}

\vspace{1cm}
\centerline{ \fbox{$\g \g \to \g \g$ and $\g \g \to \g Z$}  }
\vspace{1cm}
{\large  Contributions of operators $\O_{UB}$, $\O_{UW}$, $\ol{\O}_{UB}$
and $\ol{\O}_{UW}$} to $\g \g \to \g \g$
\bqa
F_{\lambda\lambda'\mu\mu'}&=&
-\Big\{ d^2_B+d^2{s^4_W\over c^4_W}\Big\}\
{g^2\cwd s\over{4M^2_Z}}\ \ \lambda^2\lambda^{'2}\mu^2\mu^{'2}
 \nonumber\\
&\times &\Bigm\{   {s\over{s-m^2_H}}(1+\lambda\lambda')(1+\mu\mu')
+{\ctm \over2}\ (1-\lambda\mu)\ (1-\lambda'\mu')\ {t\over{t-m^2_H}}
\nonumber\\
&-&{\ctp\over2}(1-\lambda\mu')\ (1-\lambda'\mu){u\over{u-m^2_H}})\Bigm\}
\nonumber\\
&-& \Big\{ \bar{d}^2_B+\bar{d}^2\ {s^4_W\over c^4_W}\ \Big\}
\ {g^2\cwd s\over4M^2_Z }
\lambda^2\lambda^{'2}\mu^2\mu^{'2}\nonumber\\
&\times &\Bigm\{ \ {s\over{s-m^2_H}}\ (\lambda+\lambda')(\mu+\mu')
-{\ctm\over2}\ (\mu-\lambda)\ (\mu'-\lambda')\
{t\over{t-m^2_H}}\nonumber\\
&+&  {\ctp\over2}\ (\mu'-\lambda)\ (\mu-\lambda')\
{u\over{u-m^2_H}} \Bigm\}
\eqa
Amplitudes for $\g \g \to \g Z$ are obtained by
changing $d[\bar{d}]$ into $(\cw/\sw)d[\bar{d}]$ and
$d_B[\bar{d_B}]$ into $-(\sw/ \cw)d_B[\bar{d_B}]$.

\newpage
\centerline{ \fbox{$\g \g \to W^+W^-$}}
\vspace{1cm}
{\large  Contributions of operators $\O_W$,
$\O_{UB}$, $\O_{UW}$, $\ol{\O}_{UB}$
and $\ol{\O}_{UW}$} to $\g \g \to W^+W^-$
\bqa
F_{\lambda\lambda'\mu\mu'}&=&e^2\ {\lambda_W\
s\over{M^2_W}}\ \lambda^2\lambda^{'2}\mu^2\mu^{'2}\times
\nonumber\\[0.3cm]
& \null & \Bigm\{ ( \delta_{\lambda,\lambda'}\ \delta_{\mu,-\mu'}+
\delta_{\lambda,-\lambda'}\ \delta_{\mu,\mu'}
-2\delta_{\lambda,\lambda'}\ \delta_{\mu,\mu'}\ \delta_{\lambda,-\mu}+
\nonumber\\
& \null & + {\lambda_W\ s\over{M^2_W}}\Bigm[
{(1-\cos\theta)\ (3+\cos\theta)\over16}
\delta_{\lambda',\mu}\
\delta_{\lambda,-\lambda'}\ \delta_{\mu,-\mu'}\nonumber\\
&\null & +{3-\cos^2\theta\over8}
\delta_{\lambda,\lambda'}\ \delta_{\mu,\mu'}\ \delta_{\lambda,-\mu}+
{(1+\cos\theta)(3-\cos\theta)\over16}\delta_{\lambda,\mu}\
\delta_{\lambda',\mu'}\ \delta_{\lambda,-\lambda'} \Bigm]\ \Bigm\}
- \nonumber\\
& \null & -{g^2\cwd
d_B\over{4M^2_W}}\ {s^2\over(s-m^2_H)}\ (1+\lambda\lambda')\ (1-\mu
^2)\ (1-\mu^{'2})\lambda^2\lambda^{'2} \nonumber\\
&-& {g^2\swd d\over{4M^2_W}}\ {s^2\over(s-m^2_H)}\
(1+\lambda\lambda')\ \Bigm[(1-\mu^
2)(1-\mu^{'2})+\ d\ \mu^2\mu^{'2}(1+\mu\mu')\Bigm]\
\lambda^2\lambda^{'2}\nonumber\\
&+& i\ {g^2\cwd\bar{d_B}\over{4M^2_W}}\ {s^2\over(s-m^2_H)}\
(\lambda+\lambda')\
(1-\mu^2)(1-\mu^{'2})\lambda^2\lambda^{'2}  \\
&+& i\ {g^2\swd\bar{d}\over{4M^2_W}}\ {s^2\over(s-m^2_H)}\
(\lambda+\lambda')\
\Bigm[ (1-\mu^2)\ (1-\mu^{'2})\
+\ i\ \bar{d}\ \mu^2\mu^{'2}(\mu+\mu')\Bigm] \lambda^2\lambda^{'2}
\nonumber
\eqa

\vspace{1cm}
{\large Contributions of operators $\O_{B\Phi}$ and $\O_{W\Phi}$}
to $\g \g \to W^+W^-$
\bqa
F_{+--+}&=&F_{-++-}= -e^2(f^2_B+f^2_W){\ctm s\over{32M^2_W}}
\nonumber\\
 F_{+-+-}&=& F_{-+-+}= e^2(f^2_B+f^2_W){\ctp s\over{32M^2_W}}
\nonumber\\
F_{+-00}&=&F_{-+00}=F_{++--}=F_{--++}=
e^2(f^2_B+f^2_W){ s\over{16M^2_W}}
\nonumber\\
F_{++00}&=&F_{--00}= -2e^2(f^2_B+f^2_W){ s\over{16M^2_W}}
-e^2(f_B+f_W){s\over{2M^2_W}}
\eqa

\newpage

\centerline{ \fbox{$\g \g \to ZZ$}}
\vspace{1cm}
{\large  Contributions of operators $\O_{UB}$, $\O_{UW}$, $\ol{\O}_{UB}$
and $\ol{\O}_{UW}$} to $\g \g \to ZZ$
\bqa
F_{\lambda\lambda'\mu\mu'} &= &-\ {g^2 \over{4M^2_Z}}\
[d_B+d\ {\swd\over \cwd}]\
 {s^2\over(s-m^2_H)}\ (1+\lambda\lambda')\ (1-\mu^2)\
(1-\mu^{'2})\lambda^2\lambda^{'2}
\nonumber\\
& \null & -{g^2\swd\cwd
s\over{4M^2_W}}(d^2_B+d^2)\lambda^2\lambda^{'2}\mu^2\mu^{'2}\Bigm\{
({s\over(s-m^2_H)})
(1+\lambda\lambda')\ ((1+\mu\mu')
\nonumber\\
& \null & + {\ctm\over2}(1-\lambda\mu)\ (1-\lambda'\mu')\ {t\over{t-m^2_H}}
\ \nonumber \\
& \null &
-{\ctp\over2}(1-\lambda\mu')\ (1-\lambda'\mu)\ {u\over{u-m^2_H}} \Bigm\}
\nonumber\\
 &\null &+i\ {g^2 \over{4M^2_W}}\ [\bar{d_B}\ \cwd+\bar{d}\ \swd]\
  {s^2\over(s-m^2_H)}\ (\lambda+\lambda')\lambda^2\lambda^{'2}
 (1-\mu^2)\ (1-\mu^{'2})
\nonumber\\
 &\null & -(\bar{d}^2 +\bar{d_B}^2)\cwd\swd\ {g^2 \over{4M^2_W}}\
{s^2\over(s-m^2_H)}(\mu+\mu')\mu^2\mu^{'2}\ (\lambda+\lambda')
\lambda^2\lambda^{'2}
\nonumber\\
  &\null & +{g^2\swd\cwd\over{16M^2_W}}\ [\bar{d^2_B}+\bar{d^2}]\ s^2\Bigm\{
{\ctm^2\over{t-m^2_H}}(\mu-\lambda)\ (\mu'-\lambda')
\nonumber\\
 &\null & +  {\ctp^2\over{u-m^2_H}}(\mu'-\lambda)\ (\mu-\lambda')\Bigm\}
\lambda^2\lambda^{'2}\mu^2\mu^{'2}
\eqa\\
{\large  Contributions of operators $\O_{B\Phi}$ and $\O_{W\Phi}$ }
to $\g \g \to ZZ$
\bqa
F_{+--+}&=&F_{-++-}= -g'^2\ (f^2_B+f^2_W)\ {\ctm^2
s^2\over{64M^2_W(t-m^2_H)}}
\nonumber\\
 F_{+-+-}&=& F_{-+-+}= -g'^2\ (f^2_B+f^2_W)\ {\ctp^2
s^2\over{64M^2_W(u-m^2_H)}}
\nonumber\\
 F_{++--}&=& F_{--++}= -g'^2\ (f^2_B+f^2_W)\ {s^2\over{64M^2_W}}
\Bigm\{ {\ctm^2\over{(t-m^2_H)}}+{\ctp^2\over{(u-m^2_H)}}\Bigm\}
\eqa

\newpage

\centerline{ \fbox{$\g \g \to H\ H$}  }
\vspace{1cm}
{\large  Contributions of operators $\O_{UB}$, $\O_{UW}$, $\ol{\O}_{UB}$
and $\ol{\O}_{UW}$} to $\g \g \to H\ H$
\bqa
F_{\lambda\lambda'}&=&-\lambda^2\lambda^{'2} g^2
\Bigm[ { (\cwd\ d_B^2+\swd\ d^2)\ s\over2\mwd}\ (1+3\lambda
\lambda')+
 { (\cwd\ d_B+\swd\ d)\ s\over4\mwd}\ (1+\lambda\lambda')
\Bigm]
\nonumber\\
&\null & +\lambda^2\lambda^{'2} g^2 \Bigm[ { (\cwd\
\bar{d}_B^2+\swd\ \bar{d}^2)\ s\over2\mwd}\ (3+\lambda
\lambda')+i
 { (\cwd\ \bar{d}_B+\swd\ \bar{d})\ s\over4\mwd}\
(\lambda+\lambda')
\Bigm]
\eqa

{\large  Contributions of operators $\O_{B\Phi}$, $\O_{W\Phi}$}
to $\g \g \to H\ H$
\bq
F_{--}=-2F_{-+}=-e^2(f^2_B+f^2_W)\ {s\over8\cwd\mwd}   \eq

\newpage

\centerline { {\bf Figure Captions }}\par

Fig.1 Sensitivity to the operator $\O_W$ in $\gamma\gamma \to
W^+W^-$. Transverse momentum ($p_T$) distribution $d\sigma/d p_T$
(a) at 0.5 TeV, (b) at 1 TeV.\\

Fig.2 Sensitivity to the operators $\O_{B\Phi}$ and $\O_{W\Phi}$
in $\gamma\gamma \to W^+W^-$, $ZZ$, $HH$. (a), (b), same
captions.\\

Fig.3 Sensitivity to the operators $\O_{UW}$ and $\ol{\O}_{UW}$ in
$\gamma\gamma \to W^+W^-$, $ZZ$, $\gamma Z$, $\g\g$, $HH$. (a), (b), same
captions.\\

Fig.4 Sensitivity to the operators $\O_{UB}$ and $\ol{\O}_{UB}$
in $\gamma\gamma \to W^+W^-$, $ZZ$, $\gamma Z$, $\g\g$, $HH$.
(a), (b), same captions.\\

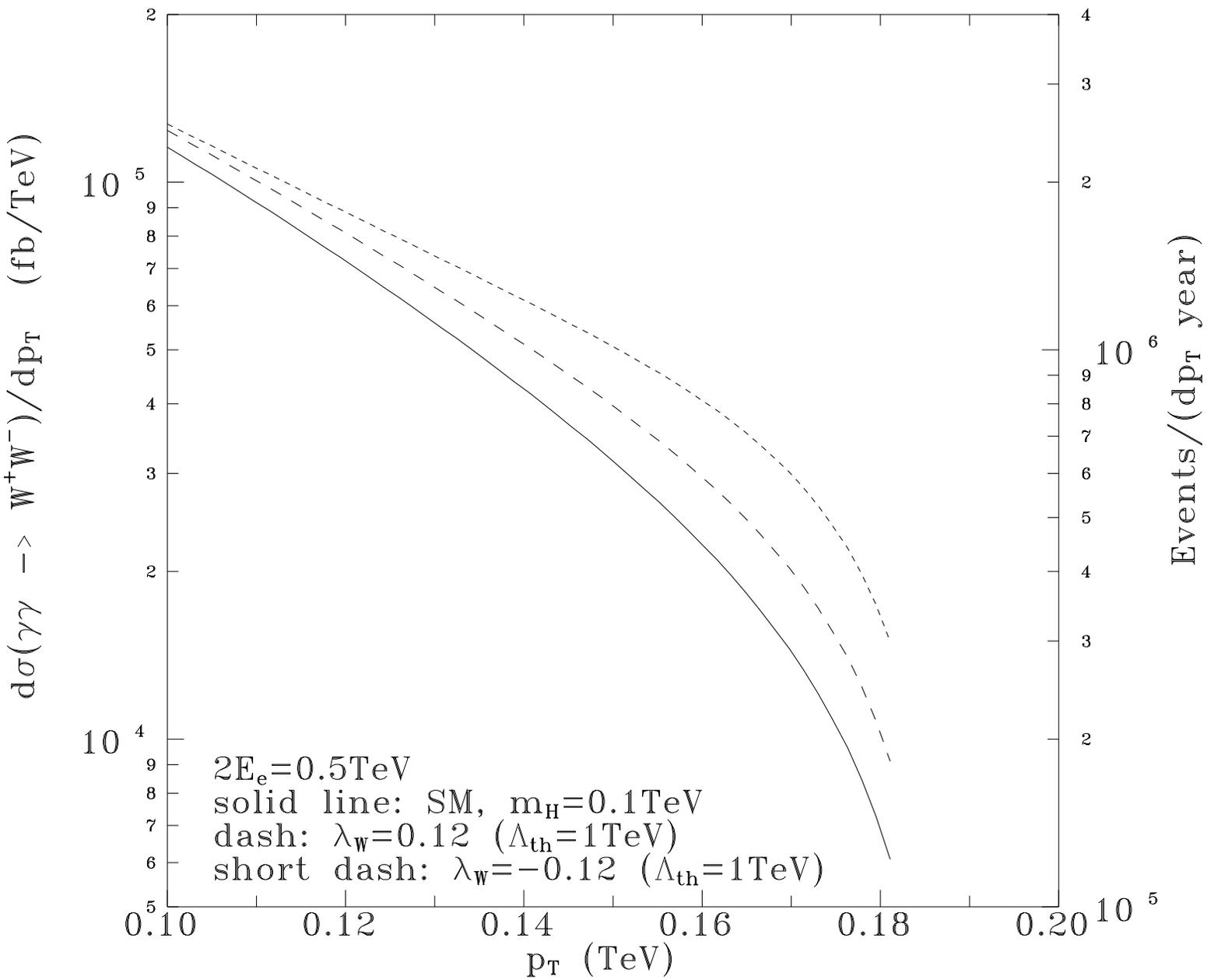
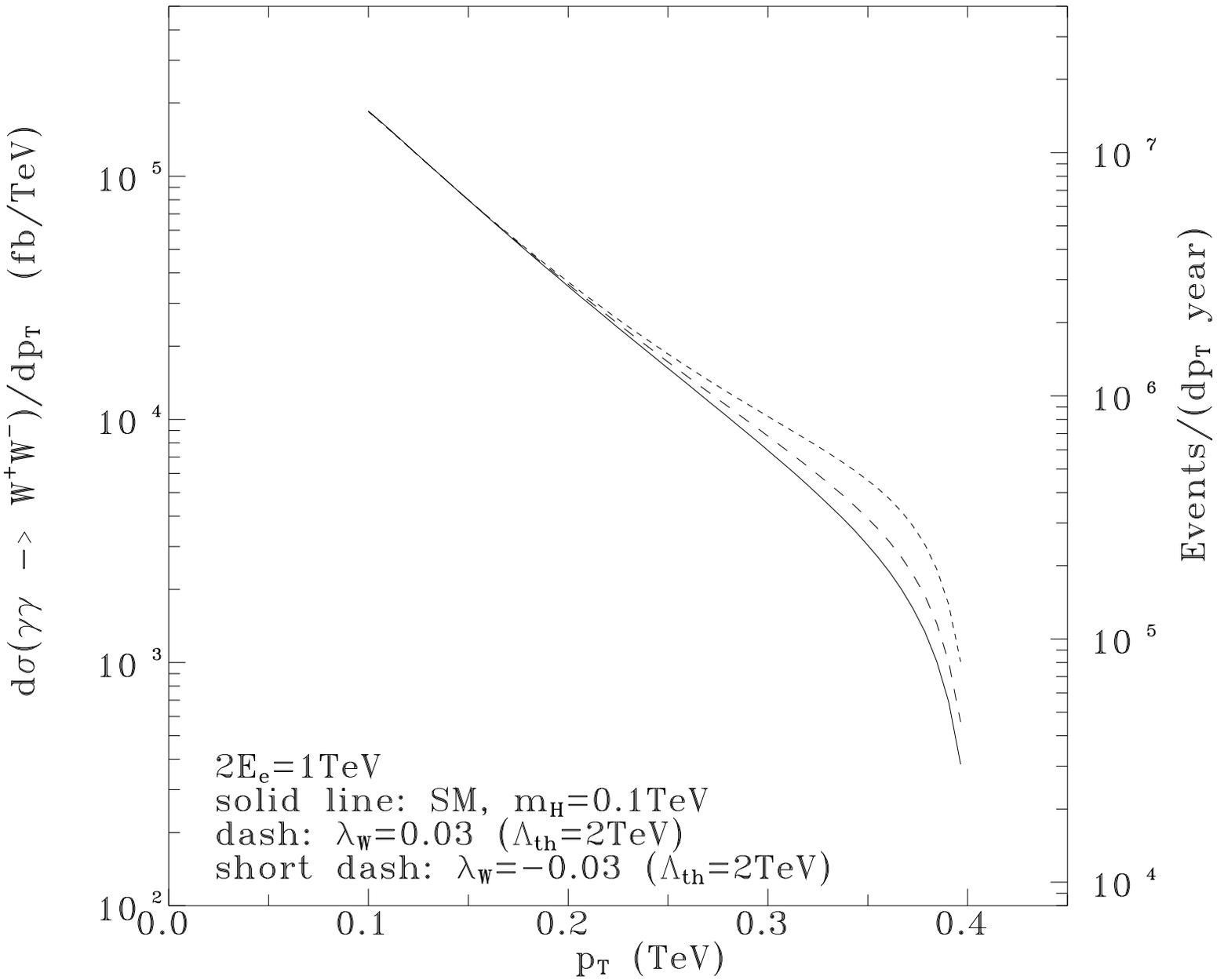
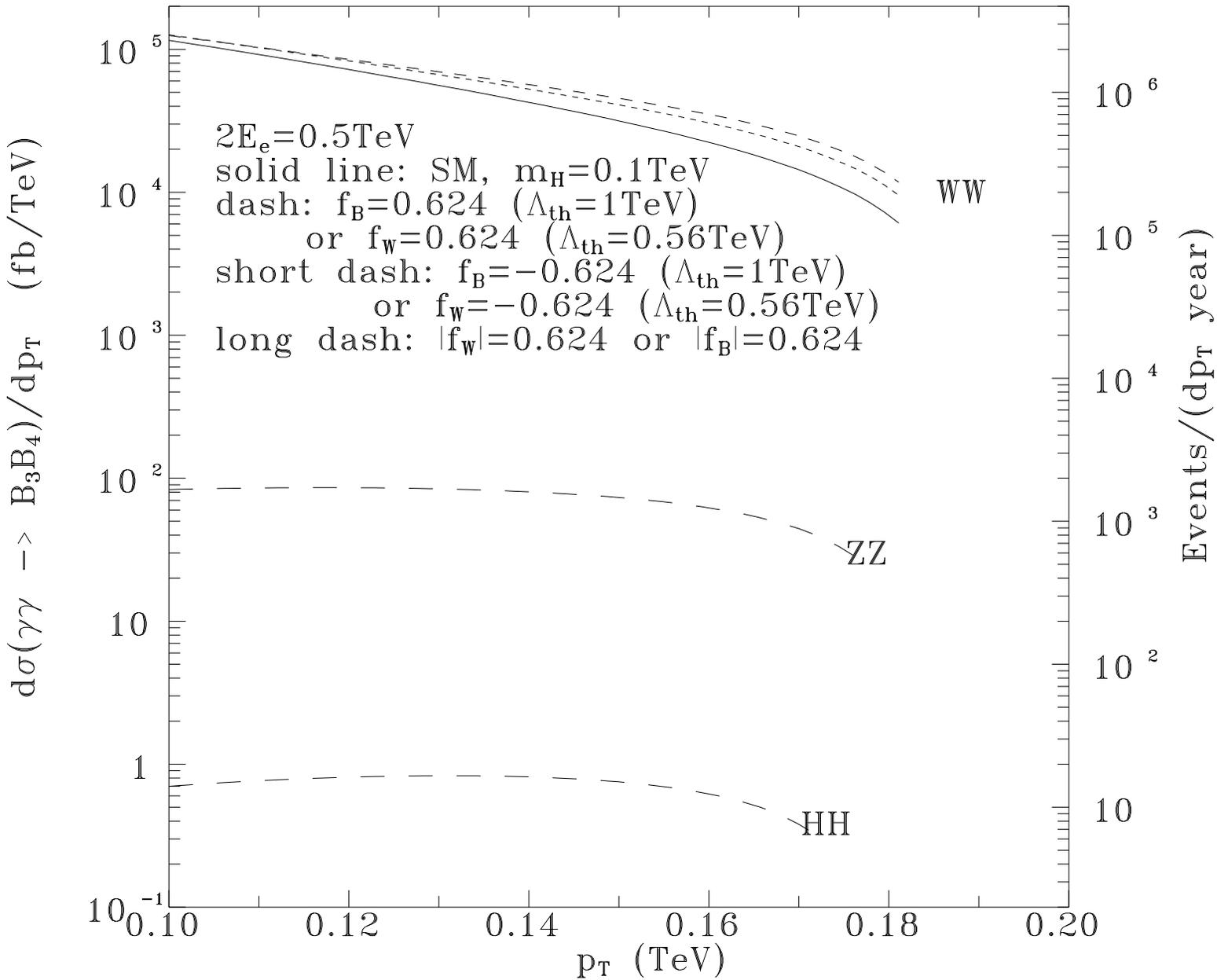
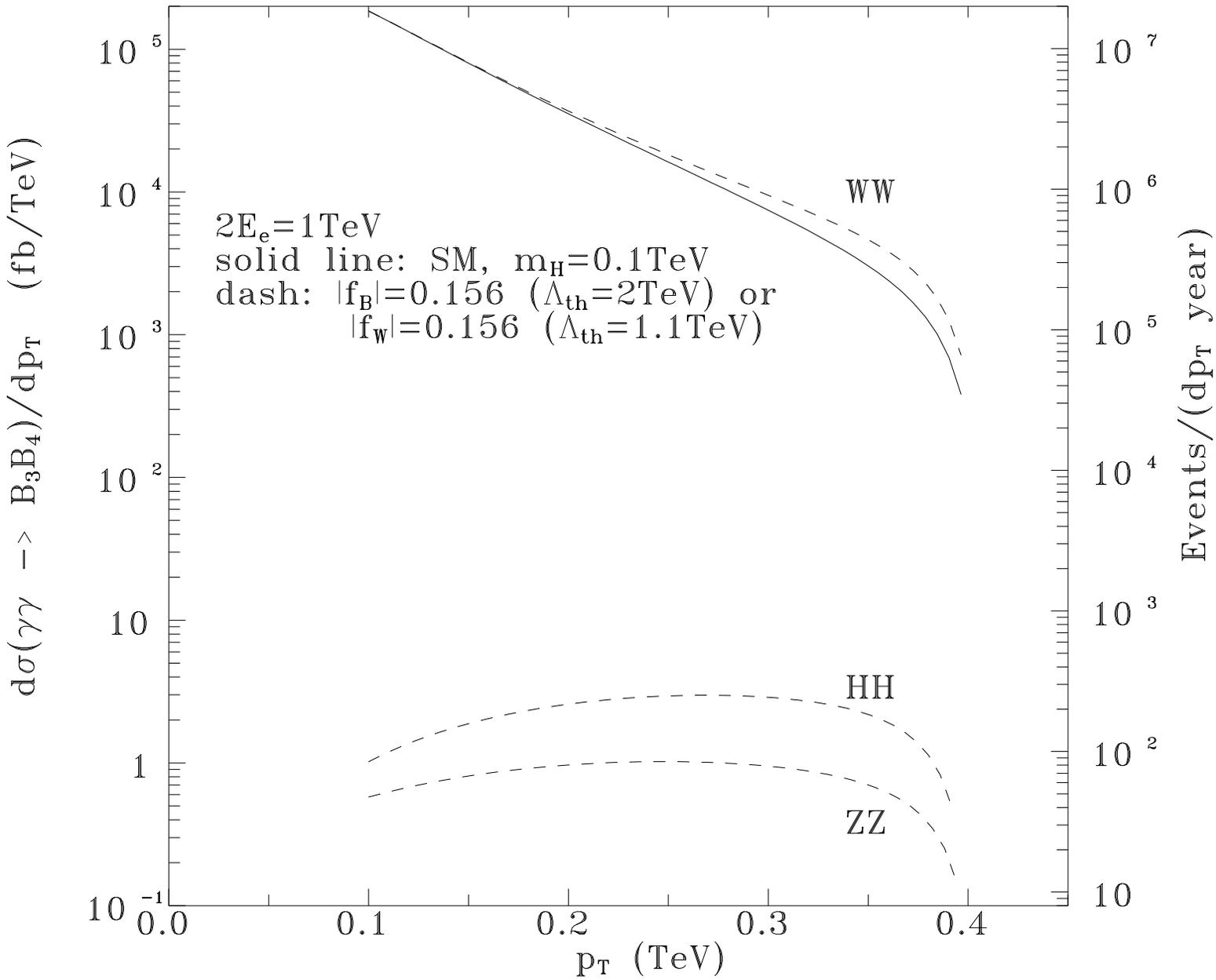
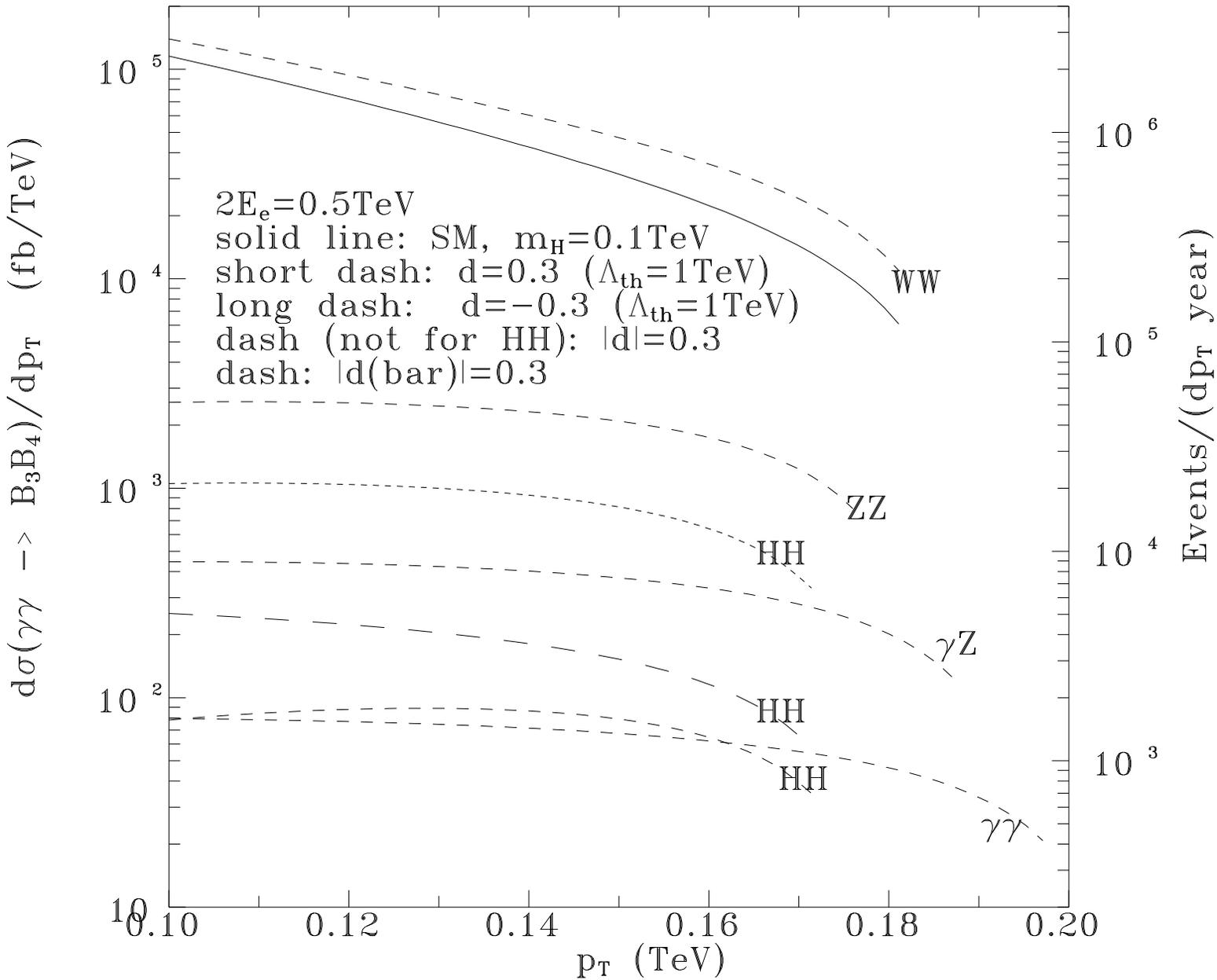
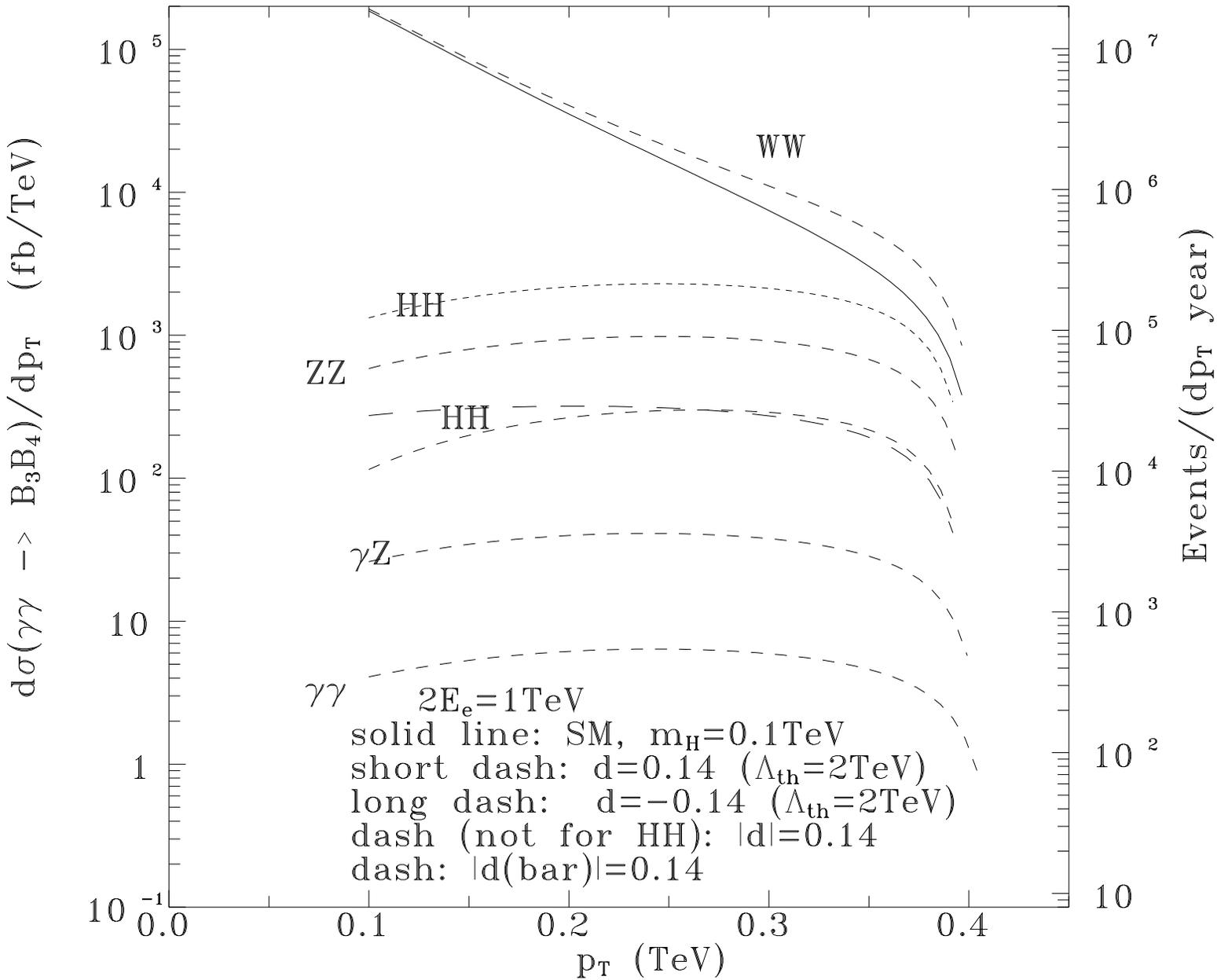
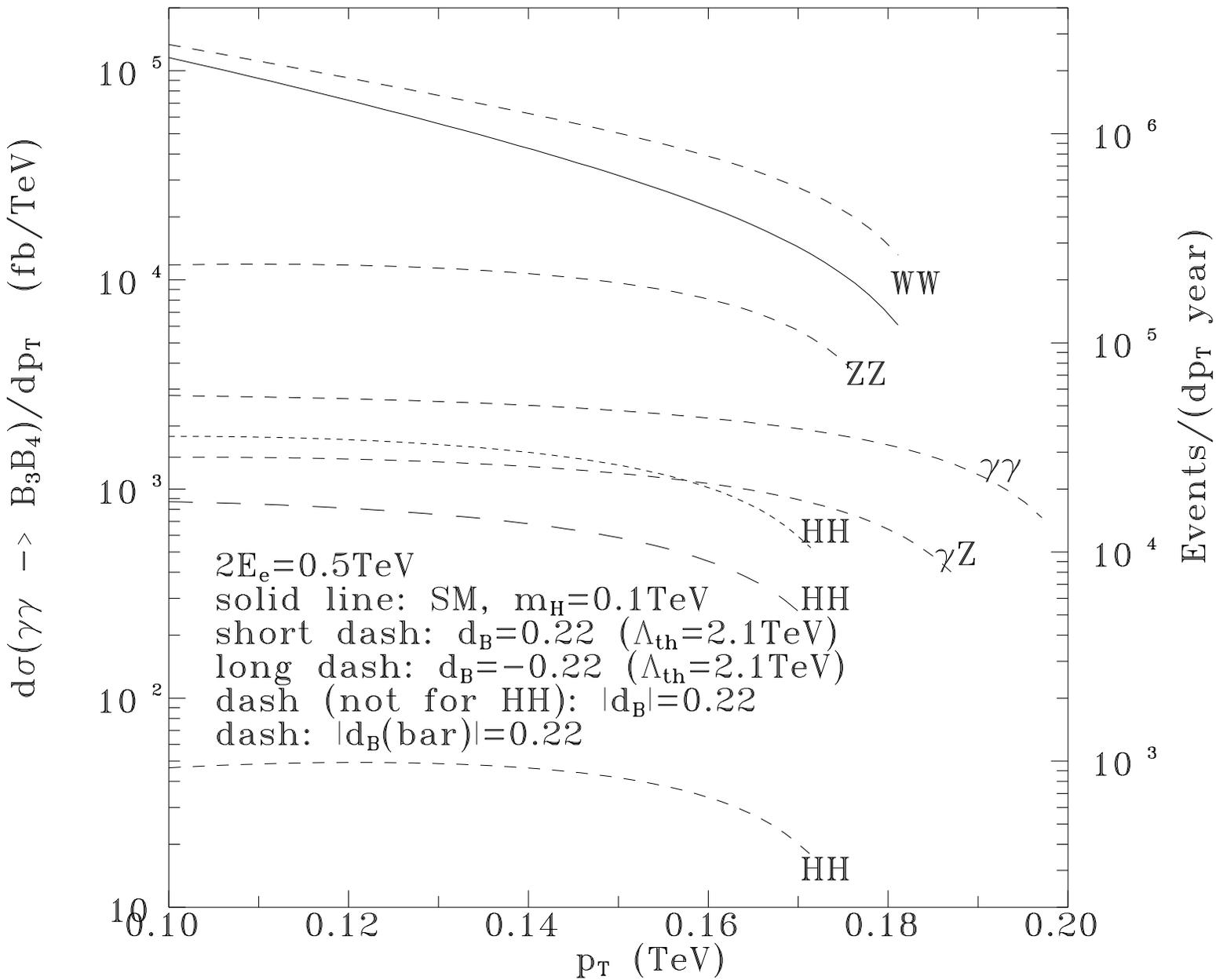
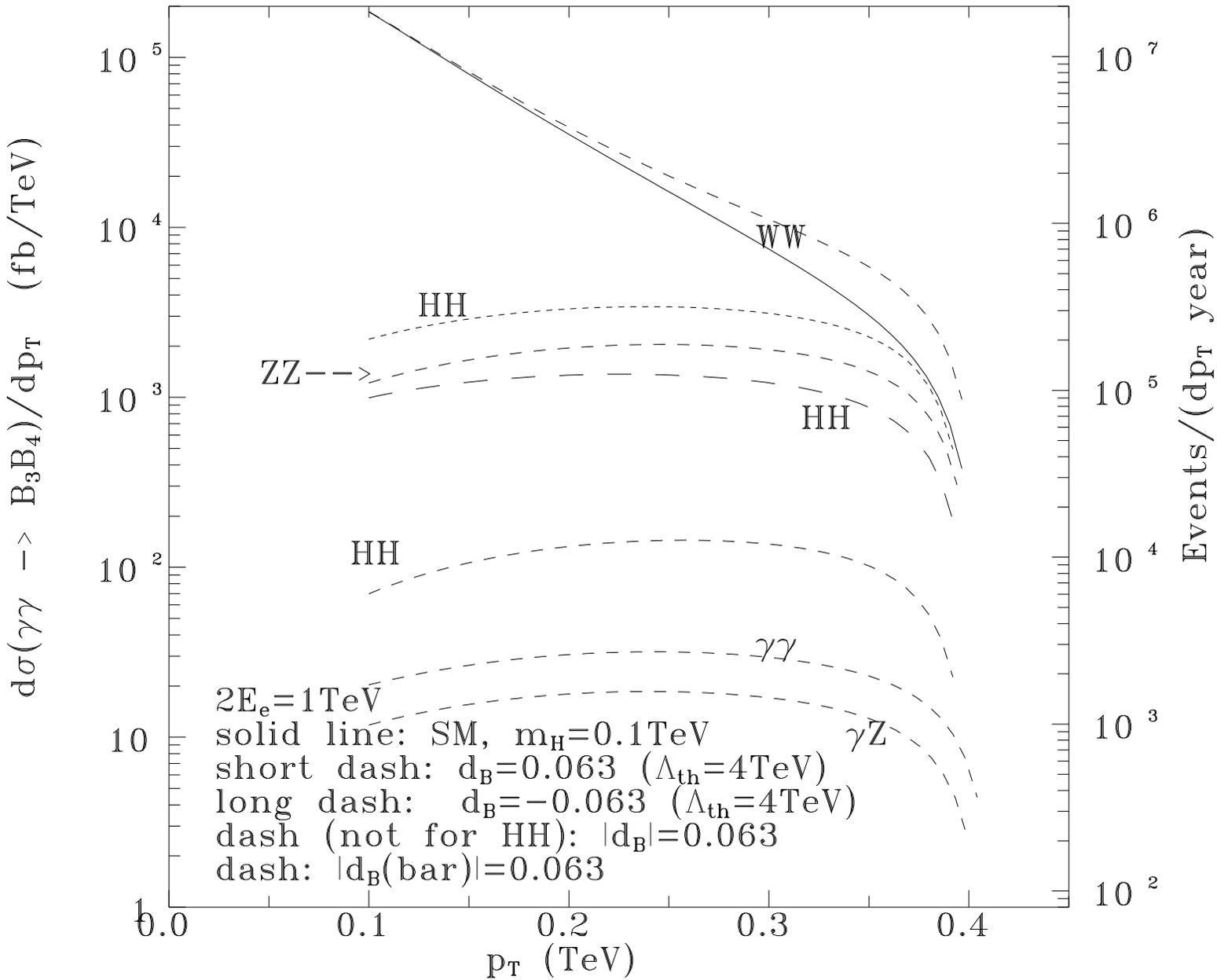
\end{document}